\documentclass[aps,twocolumn,superscriptaddress,showkeys,prb]{revtex4-2}
\AtBeginDocument{}
\usepackage[english]{babel}
\usepackage[T1]{fontenc}
\usepackage{amsmath}
\usepackage{siunitx}
\DeclareSIUnit{\molar}{M}
\DeclareSIUnit\rpm{rpm}
\DeclareSIUnit\pixel{px}
\usepackage{chemformula}
\DeclareMathOperator\Arg{Arg}
\usepackage{graphicx}
\usepackage[colorlinks=true,linkcolor=blue,urlcolor=blue,citecolor=blue]{hyperref}
\newcommand{\mi}{\mathrm{i}}
\usepackage{multirow}
\makeatletter
\renewcommand{\fnum@figure}{Figure \thefigure}
\renewcommand{\thetable}{\arabic{table}}
\makeatother
\usepackage{hyperref}
\newcommand*{\figref}[2][]{%
  \hyperref[{fig:#2}]{%
    Figure~\ref*{fig:#2}%
    \ifx\\#1\\%
    \else
      \,#1%
    \fi
  }%
}
\newcommand*{\tabref}[2][]{%
  \hyperref[{tab:#2}]{%
    Table~\ref*{tab:#2}%
    \ifx\\#1\\%
    \else
      \,#1%
    \fi
  }%
}
\newcommand*{\secref}[2][]{%
  \hyperref[{sec:#2}]{%
    Section~\ref*{sec:#2}%
    \ifx\\#1\\%
    \else
      \,#1%
    \fi
  }%
}
\newcommand*{\equref}[2][]{%
  \hyperref[{eq:#2}]{%
    Equation~\ref*{eq:#2}%
    \ifx\\#1\\%
    \else
      \,#1%
    \fi
  }%
}
\begin{document}

\author{Yonas Lebsir}
\email{lebsir@physik.uni-kiel.de}
\affiliation{Centre for Nano Optics, University of Southern Denmark, 5230 Odense, Denmark}
\affiliation{Institute for Experimental and Applied Physics, Kiel University, 24118 Kiel, Germany}

\author{Sergejs Boroviks}
\affiliation{Centre for Nano Optics, University of Southern Denmark, 5230 Odense, Denmark}
\affiliation{Nanophotonics and Metrology Laboratory (NAM), Swiss Federal Institute of Technology, Lausanne (EPFL), 1015 Lausanne, Switzerland}

\author{Martin Thomaschewski}
\affiliation{Centre for Nano Optics, University of Southern Denmark, 5230 Odense, Denmark}

\author{Sergey I. Bozhevolnyi}
\affiliation{Centre for Nano Optics, University of Southern Denmark, 5230 Odense, Denmark}
\affiliation{Danish Institute for Advanced Study, University of Southern Denmark, 5230 Odense, Denmark}

\author{Vladimir A. Zenin}
\email{zenin@mci.sdu.dk}
\affiliation{Centre for Nano Optics, University of Southern Denmark, 5230 Odense, Denmark}

\title{Ultimate limit for optical losses in gold, revealed by quantitative near-field microscopy}
\date{\today}

\begin{abstract}
We report thorough measurements of surface plasmon polaritons (SPPs) running along nearly perfect air-gold interfaces formed by atomically flat surfaces of chemically synthesized gold monocrystals. By means of amplitude- and phase-resolved near-field microscopy, we obtain their propagation length and effective mode index at visible wavelengths (532, 594, 632.8, 729, and \SI{800}{\nano\meter}). The measured values are compared with the values obtained from the dielectric functions of gold that are reported in literature. Importantly, a reported dielectric function of monocrystalline gold implies $\sim 1.5$ times shorter propagation lengths than those observed in our experiments, whereas a dielectric function reported for properly fabricated polycrystalline gold leads to SPP propagation lengths matching our results. We argue that the SPP propagation lengths measured in our experiments signify the ultimate limit of optical losses in gold, encouraging further comprehensive characterization of optical material properties of pure gold as well as other plasmonic materials.
\end{abstract}
\maketitle
\begin{figure}[h]
\centering\includegraphics{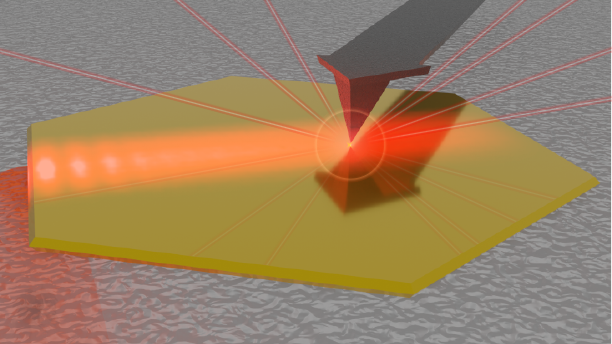}
\end{figure}
The dielectric functions of noble metals have been measured since the beginning of the plasmonic era (starting with the Wood anomaly) and the development of the Drude model \cite{Drude, Wood, Raether, Maier}. Especially in the 1980s they were intensively studied \cite{Weaver}, resulting in two highly cited works: An article by Johnson and Christy \cite{JC} and a compilation handbook by Palik \cite{Palik}. However, the rediscovery of plasmonics in the 21st century and the improvement in both fabrication facilities and methods pointed out a need to revise the optical constants of metals \cite{Babar, Blanchard, Olmon, Shalaev, Norris}. Particularly, the progress in synthesizing monocrystalline nanostructures and microstructures of noble metals \cite{metal_NC1, metal_NC2, Au_flakes, Ag_flakes} brought up a question: \textit{How much is gained by replacing a polycrystalline metal with a monocrystalline one?} It is well known that monocrystalline structures, milled with a focused ion beam, exhibit greatly improved fabrication tolerances and hence, better reproducibility, in comparison with their polycrystalline counterparts \cite{Au_flakes}. Also, the true atomical smoothness featured on chemically synthesized monocrystals can be crucial for applications with extreme light confinement \cite{Sergio_GSP, Zayats}. Yet, while the absence of electron scattering at grain boundaries generally leads to lower optical losses in a monocrystalline structure \cite{grain_scattering1, grain_scattering2}, recent spectroscopic ellipsometry measurements of the relative permittivity of gold show no significant difference between mono- and evaporated polycrystalline gold \cite{Blanchard, Olmon, Shalaev}. Likewise, in a study of plasmonic devices, the replacement of a polycrystalline gold with a monocrystalline one shows no significant performance improvements \cite{Borovik_Metasurf}. The superiority of monocrystalline gold flakes in plasmonic nanostructures is only observed in a few articles \cite{slot_wg, Shailesh, Stenger}, but without providing comprehensive measurements of the optical properties of gold. Notably, the measured dielectric function of a polycrystalline gold, fabricated with an improved recipe by McPeak et al. \cite{Norris}, is found to be superior (in terms of losses) to all previously reported dielectric functions of gold (both mono- and polycrystalline gold, see \secref{S1} in Supporting Information). Thus, it remains an open question what the permittivity of the perfect (i.e., atomically flat and monocrystalline) gold is, implying that this would then bring about an agreement between simulations and experiments with the perfect gold, especially those involving surface plasmons whose properties are extremely sensitive to the gold permittivity. Answering this question would also enable more accurate design and performance prediction of novel plasmonic devices.

In this work, we employ a scattering-type scanning near-field optical microscope (s-SNOM) to carefully map phases and amplitudes of surface plasmon-polaritons (SPPs) travelling along clean surfaces of monocrystalline gold flakes. We then accurately determine the SPP phase gradient (in the propagation direction) and propagation length, which are straightforwardly related to the gold permittivity, with the latter being directly connected with optical losses in metal. This SPP characterization is conducted at several free-space wavelengths, namely at 532, 594, 632.8, 729, and \SI{800}{\nano\meter}, allowing us to compare thus established SPP effective mode indices and propagation lengths with those obtained by using different measurements of the dielectric function of gold reported in literature. We discuss possible reasons for discrepancies and provide suggestions for further improving the measurement accuracy.

The monocrystalline Au flakes were synthesized on a glass substrate, using a modified Brust-Schiffrin method for colloidal synthesis of gold particles \cite{Brust-Schiffrin}, as described by Radha and Kulkarni \cite{Radha} (see \secref{S2} in Supporting Information). This process is known to yield large and high-aspect-ratio flakes. Specifically, the grown flakes were up to \SI{\sim 100}{\micro\meter} in lateral size and 0.1 to \SI{2}{\micro\meter} in thickness. Two larger and visually defect-free flakes were selected for further SPP excitation and near-field measurements (see \figref[a,b]{1} and \secref{S3}, \secref{S4} in Supporting Information). Note that gold atoms are packed in a face-centered cubic (FCC) lattice, resulting in non-rectangular, tapered-edge profile of the flake (see side facets in the inset of \figref[b]{1}, discussion concerning the geometry of the flake edges and their optical properties can be found in a recent work \cite{Borovik_flakeSEM}).

\begin{figure}
\centering\includegraphics{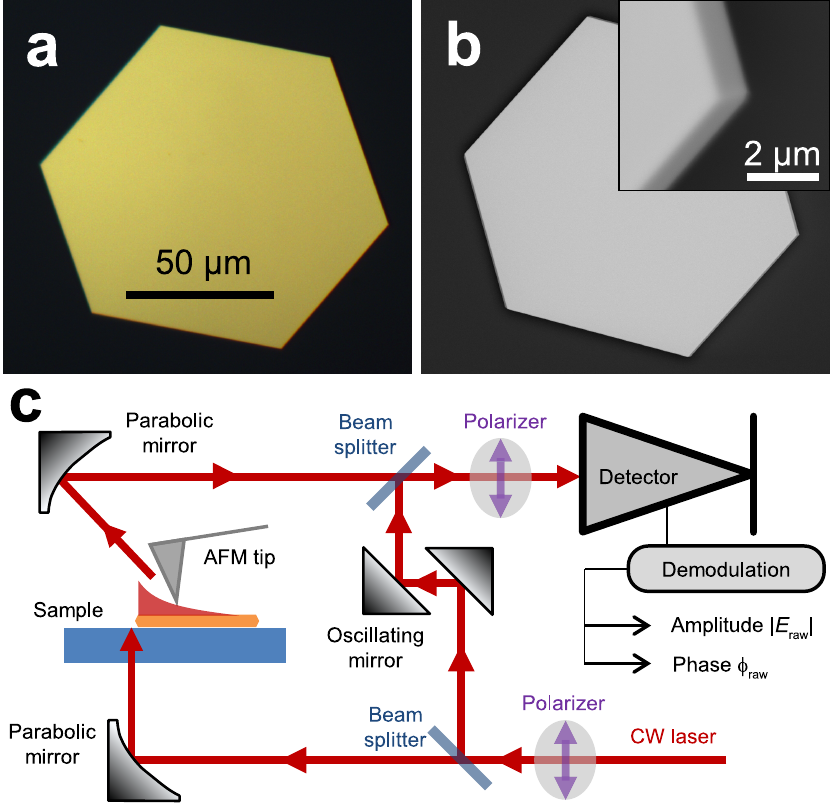}
  \caption{(a) Bright-field and (b) SEM images of the sample with a ca. \SI{1.4}{\micro\meter} thick gold flake on the glass substrate. Inset in (b) shows close-up of the flake corner, demonstrating protruding side facets of the flake. (c) Setup layout of s-SNOM.}
  \label{fig:1}
\end{figure}

Synthesized monocrystalline gold flakes are too small for a variable angle spectroscopic ellipsometer\textemdash currently the most widely used tool to characterize the dielectric function of a material. Another option is to fabricate sets of in- and out-couplers on the flake, each with different separation, and to measure SPP transmission in a conventional far-field setup. However, this requires precise control (reproducibility) of the coupling efficiency, and SPP beam divergence should be properly considered. Additionally, this far-field method is not suitable for short wavelengths, where SPP propagation length (and thus the largest separation between couplers) is of the order of the illumination beam size. Therefore, the most suitable setup is an amplitude- and phase-resolved s-SNOM, operating in  transmission mode (see \figref[c]{1} and \hyperref[sec:Methods]{Methods}). Here, the bottom parabolic mirror focuses the laser beam at the flake edge to excite SPPs. The SPP near field is then scattered by a SNOM probe and mapped by synchronously raster-scanning both the bottom parabolic mirror and the sample. Analysis of our measurements unveiled small deviations from this synchronicity as a source of noticeable errors. These deviations are due to the different drive techniques used in the positioner of the bottom parabolic mirror (a Stick-Slip stage from SmarAct that combines piezo and friction elements) and the sample stage (continuous piezo movement). With additional measurements, we estimated the lateral mismatch between sample and bottom parabolic mirror to be random and on the order of \SI{1}{\micro\meter} per \SI{100}{\micro\meter} lateral displacement (but it was not accumulating during the scan\textemdash see \secref{S5} in Supporting Information).

\begin{figure*}
\centering\includegraphics{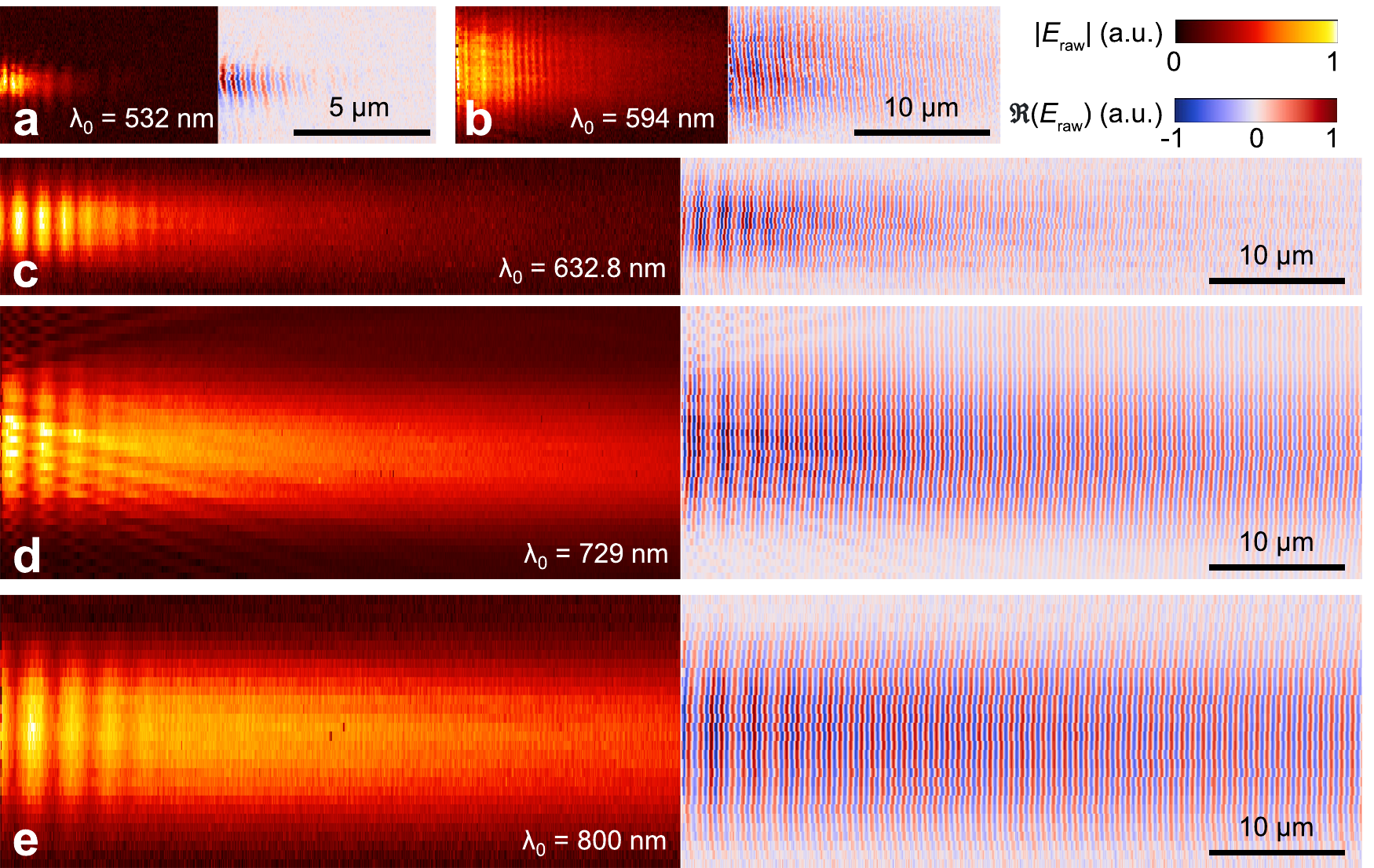}
  \caption{Amplitude, $|E_{\mathrm{raw}}|$, and the real part, $\Re(E_{\mathrm{raw}})$, of the measured near-field maps, demonstrating the propagation of SPPs, excited at the flake edge (placed \SI{\sim 500}{\nano\meter} beyond the left edge of the image) with a laser of the free-space wavelength of (a) \SI{532}{\nano\meter}, (b) \SI{594}{\nano\meter}, (c) \SI{632.8}{\nano\meter}, (d) \SI{729}{\nano\meter}, and (e) \SI{800}{\nano\meter}. Note the twice as small lateral scale bar in (a), compared with the others.}
  \label{fig:2}
\end{figure*}

The as-recorded near-field maps, represented in terms of the electric field amplitude, $|E_{\mathrm{raw}}|$, and its real part, $\Re(E_{\mathrm{raw}})$, are shown in \figref{2}. To allow for a high scanning speed without damaging the tip, the flake edge was positioned approximately \SI{0.5}{\micro\meter} away from the left end of the scan. The incident laser beam was centered at the selected flake edge, with the polarization being perpendicular to the edge. Although not as efficient as a grating coupler, illumination of the edge allows for the excitation of a SPP propagating perpendicularly to the flake edge. To have a negligible divergence of the SPP beam, the diameter of the freely propagating laser beam was adjusted to produce a large enough spot size after being focused with the bottom parabolic mirror (see \secref{S6} in Supporting Information).

One can notice that additionally to the expected decaying SPP beam, the near-field maps demonstrate interference fringes near the excitation edge (\figref[]{2}). Although in both the direct and Fourier space, this additional contribution looks like another quickly decaying mode with an effective mode index of around 1.3, we suppose that it is due to the reflection of the incident light by a tip facet (see \secref{S8} in Supporting Information). In order to avoid any influence of this contribution, we applied a Fourier-filtering procedure based on the EDFT algorithm \cite{EDFT} ($E_{\mathrm{raw}} \rightarrow E_{\mathrm{f}}$), and further neglected the data from the first \SI{20}{\percent} of propagation distance in the SPP fitting procedure. Finally, to exclude the influence of this artificial mode and any other artefacts of the setup, we measured near-field maps of the SPPs in three different cantilever-flake configurations (see \secref{S9} in Supporting Information) and used two different flakes.

\begin{figure}
\centering\includegraphics{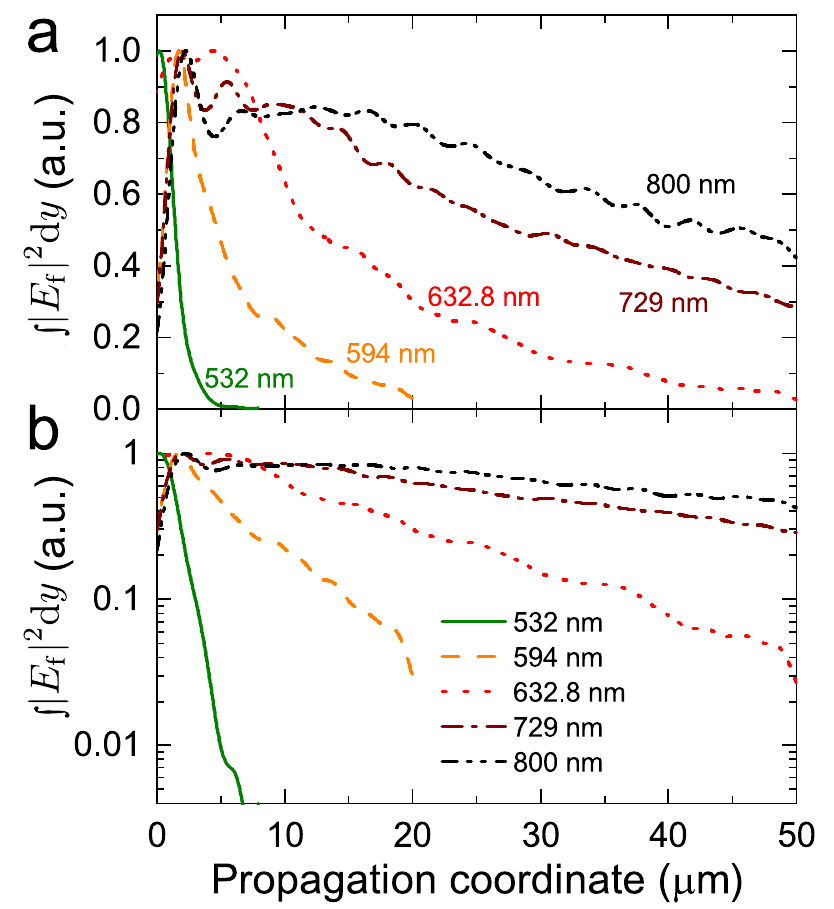}
  \caption{Evolution of Fourier-filtered SPP intensity (integrated $|E_{\mathrm{f}}|^2$ over each cross-section), plotted in (a) linear and (b) logarithmic scale for different excitation wavelengths.}
  \label{fig:3}
\end{figure}

To account for the SPP beam divergence (although being small, see estimation in \secref{S6} in Supporting Information), we integrated the square of the Fourier-filtered field amplitude, $|E_{\mathrm{f}}|^2$, for every cross-section, which revealed the SPP decay traces (\figref[]{3}). By fitting these traces with a single decaying exponential we could directly determine both the propagation length, $L_{\mathrm{p}}$, and the imaginary part of the SPP effective mode index, $\Im(N_{\mathrm{SPP}})$ (\figref[]{4}; see \hyperref[sec:Methods]{Methods}). The real part of the effective mode index, $\Re(N_{\mathrm{SPP}})$, was obtained by evaluating the average gradient of the Fourier-filtered field phase along the SPP-propagation direction. Errors were estimated by considering both the errors of the individual fits and variations between different measurements (see \hyperref[sec:Methods]{Methods}).

\begin{figure}
\centering\includegraphics{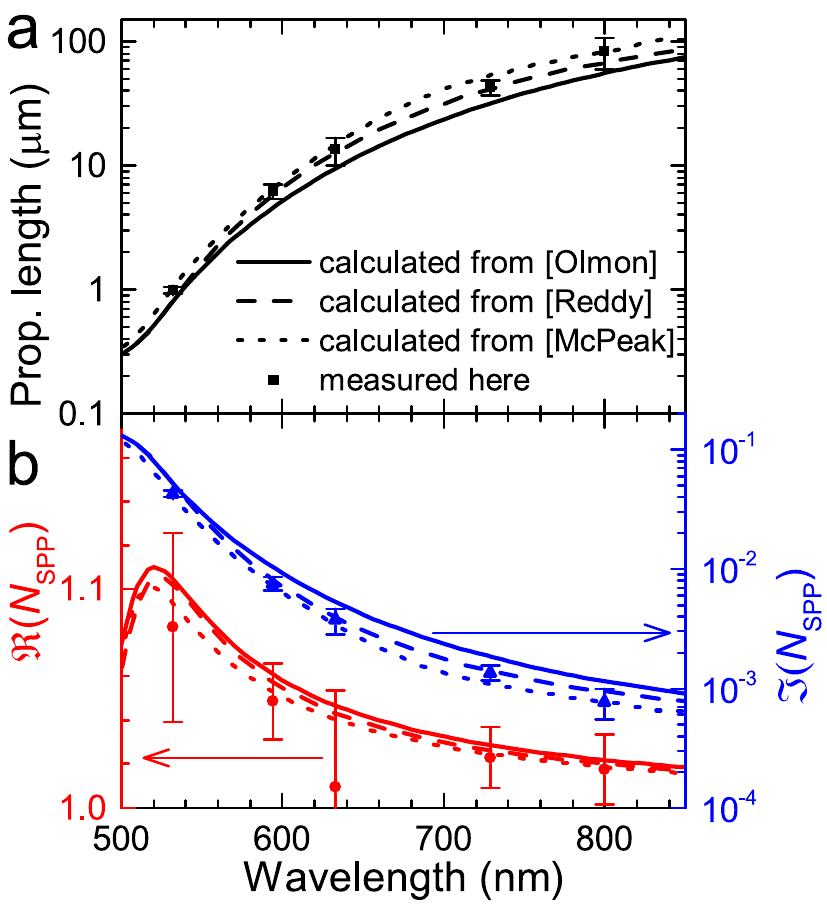}
  \caption{(a) Real $\Re(N_{\mathrm{SPP}})$ and imaginary $\Im(N_{\mathrm{SPP}})$ part of the SPP effective mode index, and (b) SPP propagation length, calculated from the measured near-field maps (dots) and compared with calculations, using gold optical properties reported by Olmon et al. \cite{Olmon}, Reddy et al. \cite{Shalaev}, and McPeak et al. \cite{Norris}.}
  \label{fig:4}
\end{figure}

Then we compared our measurements to analytically predicted values of the SPP effective mode index \cite{Raether, Maier}, $N_{\mathrm{SPP}}=\sqrt{\varepsilon_{\mathrm{Au}}/(\varepsilon_{\mathrm{Au}}+1)}$, and to the corresponding propagation lengths (\figref[]{4} and \tabref{1}). First, we took the dielectric permittivity of gold, $\varepsilon_{\mathrm{Au}}$, from published measurements for monocrystalline gold by Olmon et al. \cite{Olmon}, which seems to be the most common reference for describing optical properties of synthesized monocrystalline gold nanostructures and flakes in experimental papers. Surprisingly, the calculated propagation lengths imply about 1.5 times higher propagation losses than our measurements. A possible explanation is a considerable roughness of the used sample\textemdash a mechanically polished monocrystalline gold with a root-mean-square (RMS) roughness of \SI{1.12}{\nano\meter}. In this case, scattering by surface roughness might become dominant, reducing and eventually eliminating the benefits of using defect-free bulk material.

Next, we used the data from Reddy et al. \cite{Shalaev} (measured at room temperature prior to annealing), which implies just slightly larger losses than measured in our study. Here, the used sample seems being potentially better: it was a \SI{200}{\nano\meter} thin monocrystalline gold film, epitaxially grown on a mica substrate (purchased from Phasis Sàrl). No roughness measurements are shown, but the information from Phasis Sàrl suggests atomically smooth gold on each mica terrasse, with the average lateral size of the terrasses being \SI{\sim 0.5}{\micro\meter}, and the terrasse step height being \SI{\sim 1}{\nano\meter}.

Finally, we considered the data for the currently best-quality polycrystalline gold by McPeak et al. \cite{Norris}, which implied the lowest losses and the best agreement with our measurements.  Importantly, the high deposition rate was used here to increase the grain size (and thus to reduce scattering at grain boundaries), while the template-stripping technique provided a smooth surface with RMS roughness of \SI{0.3}{\nano\meter}. Therefore we believe that the data reported by McPeak et al. \cite{Norris} currently is the most appropriate for chemically synthesized atomically-flat monocrystalline gold nanostructures.

In principle, one can also calculate the dielectric constant of gold directly from the measured complex-valued SPP effective mode index. However, to get reasonable values, the accuracy of near-field measurements should be improved. The large error in the real part of the effective mode index is, to our opinion, due to the small misalignment of the incident angle and the deviations from synchronicity between the bottom parabolic mirror and sample movements. Besides, there is a large relative error in the imaginary part of the effective mode index at longer wavelength (although it is nearly the same on an absolute scale, see \tabref{1}). This is due to the scan length, which in those cases is relatively short compared to the longer SPP propagation lengths. A limited scanning speed was applied to avoid tip degradation during scanning, and thus a limited scan length was necessary to keep a reasonable scan time. To overcome this limitation, one can shorten the SPP wavelength and propagation length by coating the gold surface with a thin high-refractive index dielectric layer. Alternatively, excitation of SPPs at the gold-substrate interface is also possible, however in this case, the gold thickness should be on the order of \SI{50}{\nano\meter} or below, to allow evanescent tails of gold-substrate SPPs to be probed by SNOM. Either way it is critical to have a high-quality dielectric (smooth, ideally monocrystalline) without defects, and without air or contamination voids between the gold and dielectric, to solely measure absorption related losses in the monocrystalline gold. Lastly, one can try applying imaging ellipsometry instead, which has a spatial resolution down to \SI{\sim 1}{\micro\meter}, to measure the dielectric constants of gold monocrystals directly. However, this method is associated with other difficulties: due to gold’s high reflectance, fitting the ellipsometry data is nontrivial and ambiguous. We also foresee a demand for revising the dielectric constant of synthesized monocrystalline silver\textemdash in this case the difference with its polycrystalline phase might be dramatic due to the chemical reactivity of silver.

\begin{table*}
    \centering
    \begin{tabular}{l|l|l|l|l|l}
$\lambda_0$ & \SI{532}{\nano\meter} & \SI{594}{\nano\meter} & \SI{632.8}{\nano\meter} & \SI{729}{\nano\meter} & \SI{800}{\nano\meter} \\\hline
$\varepsilon_{\mathrm{Au}}$ [Olmon] & $-4.578 + 1.915\mi$ & $-8.712+1.352\mi$ & $-11.3+1.2\mi$ & $-18.116 + 1.11\mi$ & $-23.6 + 1.203\mi$ \\
$\varepsilon_{\mathrm{Au}}$ [Reddy] & $-4.654 + 1.971\mi$ & $-9.264+1.206\mi$ & $-12.151+1.039\mi$ & $-19.663 + 1.005\mi$ & $-25.659 + 1.17\mi$ \\
$\varepsilon_{\mathrm{Au}}$ [McPeak] & $-5.235 + 1.974\mi$ & $-10.01+1.265\mi$ & $-12.999+1.034\mi$ & $-21.024 + 0.89\mi$ & $-27.276 + 1.089\mi$ \\\hline
$N_{\mathrm{SPP}}$ [Olmon] & $1.105 + 5.26\cdot 10^{-2}\mi$ & $1.061+1.04\cdot 10^{-2}\mi$ & $1.047+5.33\cdot 10^{-3}\mi$ & $1.029 + 1.83\cdot 10^{-3}\mi$ & $1.022 + 1.15\cdot 10^{-3}\mi$ \\
$N_{\mathrm{SPP}}$ [Reddy] & $1.102 + 5.19\cdot 10^{-2}\mi$ & $1.058+0.82\cdot 10^{-2}\mi$ & $1.044+3.97\cdot 10^{-3}\mi$ & $1.026 + 1.4\cdot 10^{-3}\mi$ & $1.02 + 0.94\cdot 10^{-3}\mi$ \\
$N_{\mathrm{SPP}}$ [McPeak] & $1.093 + 4.13\cdot 10^{-2}\mi$ & $1.053+0.73\cdot 10^{-2}\mi$ & $1.041+3.42\cdot 10^{-3}\mi$ & $1.025 + 1.08\cdot 10^{-3}\mi$ & $1.019 + 0.77\cdot 10^{-3}\mi$ \\
\multirow{2}{*}{$\boldsymbol{N_{\mathrm{SPP}}}$ \textbf{[this work]}} & $\mathbf{(1.08\pm 0.04)\,+}$ & $\mathbf{(1.05\pm 0.02)\, +}$ & $\mathbf{(1.01\pm 0.04) \,+}$ & $\mathbf{(1.02\pm 0.02) \,+}$ & $\mathbf{(1.02\pm 0.02) \,+}$ \\
& $\mathbf{(4.3\pm 0.3)\cdot 10^{-2}\mi}$ & $\mathbf{(0.8\pm 0.1)\cdot 10^{-2}\mi}$ & $\mathbf{(3.8\pm 0.9)\cdot 10^{-3}\mi}$ & $\mathbf{(1.4\pm 0.2)\cdot 10^{-3}\mi}$ & $\mathbf{(0.8\pm 0.2)\cdot 10^{-3}\mi}$ \\\hline
$L_\mathrm{p}$ [Olmon] & \SI{0.8}{\micro\meter} & \SI{4.6}{\micro\meter} & \SI{9.4}{\micro\meter} & \SI{31.6}{\micro\meter} & \SI{55.4}{\micro\meter} \\
$L_\mathrm{p}$ [Reddy] & \SI{0.82}{\micro\meter} & \SI{5.8}{\micro\meter} & \SI{12.7}{\micro\meter} & \SI{41.4}{\micro\meter} & \SI{67.6}{\micro\meter} \\
$L_\mathrm{p}$ [McPeak] & \SI{1.02}{\micro\meter} & \SI{6.51}{\micro\meter} & \SI{14.7}{\micro\meter} & \SI{53.6}{\micro\meter} & \SI{82.4}{\micro\meter} \\
$\boldsymbol{L_\mathrm{p}}$ \textbf{[this work]} & $\mathbf{(0.99\pm 0.06)}$\,\textbf{µm} & $\mathbf{(6.2\pm 0.8)}$\,\textbf{µm} & $\mathbf{(13\pm 3)}$\,\textbf{µm} & $\mathbf{(43\pm 6)}$\,\textbf{µm} & $\mathbf{(80\pm 24)}$\,\textbf{µm} \\
\end{tabular}
\caption{Gold permittivity data reported by Olmon et al. \cite{Olmon}, Reddy et al. \cite{Shalaev}, and McPeak et al. \cite{Norris}, and the calculated SPP effective mode indices and propagation lengths, in comparison to our measurements (bold) at five different wavelengths (top row).}
\label{tab:1}
\end{table*}

In summary, we have investigated the propagation of SPPs on the surface of chemically synthesized gold crystals, using amplitude- and phase-resolved s-SNOM at visible frequencies. We analyzed the near-field maps to extract the SPP propagation properties\textemdash effective mode indices and propagation lengths\textemdash which are directly proportional to the optical losses in metal. By comparing the measured propagation lengths to the analytical calculations with the commonly used permittivity data of monocrystalline gold, we found a noticeable discrepancy, which we attribute to the surface roughness of the gold samples used in these works. Interestingly, the best agreement for our measurements was found with the dielectric function of a polycrystalline gold, as reported by McPeak et al. \cite{Norris}, where extra efforts were made to reduce roughness (by template stripping) and scattering at grain boundaries (by using a high deposition rate). Our work provides an answer to the question of what the most accurate permittivity data for chemically synthesized gold nanostructures is. We also envision further\textemdash detailed and thorough\textemdash study of the dielectric constants of the most common plasmonic metals (namely gold and silver) in their perfect state (atomically flat and monocrystalline) and broader application of this perfect material.

{\bf Note.} During the final stage of the preparation of this manuscript, a similar, recent preprint was brought to our attention \cite{Stenger_arxiv}. There, the authors use similar gold monocrystals to study SPPs excited with a \SI{632.8}{\nano\meter} He-Ne laser, and they employ the same s-SNOM, but in reflection mode, where the top parabolic mirror is used for both illumination and collection of scattering from the tip. This makes the near-field maps less intuitive to understand, and data processing is more sophisticated, but at the end, the authors arrive at the same conclusion\textemdash that the SPP propagation losses correspond to the theoretical value with McPeak \cite{Norris} gold permittivity.

\section*{Methods}\label{sec:Methods}

{\bf Near-field measurements.}
Our s-SNOM (NeaSpec GmbH) is based on an atomic force microscope (AFM) operating in amplitude-modulated tapping mode. The AFM tip (a platinum-coated Si probe, Arrow NCPt from NanoWorld) scatters the near field of propagating SPPs, which is then collected with the upper parabolic mirror. However, the bulk scattering of the incident light by the sample itself and the tip is also collected (usually referred to as background). The near-field contribution is, unlike the background, highly dependent on the tip-sample distance, and therefore the tapping mode allows for efficient background filtering by demodulating the detected signal at high-order harmonics of the tapping frequency (\SI{\sim 250}{\kilo\hertz}). Since the signal-to-noise ratio in our measurements was high enough, we used the near-field maps recorded at the 4th harmonic for our SPP propagation analysis (although measurements at the 3rd harmonic yielded similar results, see comparison in \secref{S7} in Supporting Information). In addition, the setup features a reference arm with an oscillating mirror (frequency \SI{\sim 300}{\hertz}) for interferometric pseudoheterodyne detection \cite{pseudoheterodyne}, to simultaneously measure both amplitude and phase of the near field. Our CW laser sources were a second-harmonic Nd:YAG \SI{532}{\nano\meter} laser (\SI{60}{\milli\watt}, WiTec), a \SI{594}{\nano\meter} semiconductor laser (\SI{70}{\milli\watt}, Coherent OBIS LX), a \SI{632.8}{\nano\meter} HeNe laser (\SI{40}{\milli\watt}, Melles Griot), and a tunable Ti:sapphire laser (0.5--\SI{1}{\watt}, Spectra Physics 3900S), operated at 729 and \SI{800}{\nano\meter}. The Nd:YAG and Ti:sapphire lasers were pre-coupled into fiber and out-coupled with a short focal length reflective collimator (Thorlabs RC02FC-P01) to produce a reasonably large beam spot at the sample (see estimated spot size in \secref{S6} in Supporting Information). Finally, all lasers were attenuated to about \SI{40}{\milli\watt} before entering the SNOM setup.

{\bf Processing of near-field data.}
First, the raw near-field data of a given scan was transformed with the extended discrete Fourier transform (EDFT \cite{EDFT}) along the propagation direction, then filtered with a rectangular apodization window around the SPP peak (spanning from 0.8 to 1.2 of the effective mode index for all wavelengths except for \SI{532}{\nano\meter}, where the window width was increased to span from 0.8 to 1.3). A comparison between raw and filtered data, as well as the Fourier spectrum, can be seen in \secref{S7} in the Supporting Information. Then, to find the decay trace (\figref[]{3}), the amplitude of the filtered near field, $|E_{\mathrm{f}}|$, was squared and integrated for each cross-section ($\int |E_{\mathrm{f}}|^2 dy$). By robust fitting of the decay trace with a single decaying exponential (i.e., $\exp{\left[-2k_0\Im(N_{\mathrm{SPP}})x\right]}$), we find the imaginary part of the effective mode index, $\Im(N_{\mathrm{SPP}})$, and the propagation length $L_{\mathrm{p}}=\frac{\lambda_0}{4\pi\Im(N_{\mathrm{SPP}})}$ (see \secref{S6} in Supporting Information for further explanation). Errors for the respective average values for all scans at each wavelength, as shown in \tabref{1}, were estimated from the variation in results that were extracted for different measurement configurations (using different flakes and different cantilever-flake configurations, see \secref{S9} in Supporting Information). For the real part of the effective mode index, $\Re(N_{\mathrm{SPP}})$, the phase gradient of the filtered near field along the propagation direction (i.e., $d\Arg{(E_{\mathrm{f}})}/dx$) was calculated for all image points and averaged over, using a weighted median function with $|E_{\mathrm{f}}|$ as a weight. Here, the absolute variance is used to estimate the error of the determined $\Re(N_{\mathrm{SPP}})$ value for each scan. Then, the final errors for the respective values at the different wavelengths, as shown in \tabref{1}, were estimated as the sum of the average absolute variance and the variation in results for different cantilever-flake configurations. Note that calculating the average phase gradient is more accurate than a simple determination from the peak position in the Fourier spectrum (see \secref{S8} in Supporting Information), because it overcomes the resolution limit of Fourier transform (which is due to the limited scan length).

\textbf{Associated content}
Supporting information: Plots of permittivity data reported in previous works; fabrication recipe and detailed description of the measured gold flakes; measurements of the scanning mismatch between the focused laser spot and the sample; description of the 2D SPP Gaussian beam, laser alignment and beam parameters estimated for the measurements at the different wavelengths; comparison of measurements at the 3rd and 4th harmonics of the near field; analysis of visible interference fringes in the recorded near-field maps; plots of decay traces for all cantilever-flake configurations and fitting intervals.

\textbf{Acknowledgments}
The authors acknowledge financial support from VILLUM FONDEN (grants 16498 and 40707). The authors are grateful to N. Asger Mortensen for his support and stimulating discussions.

\textbf{Notes}
The authors declare no competing financial interest. The data that support the findings of this study are available from the corresponding authors upon reasonable request.

\bibliography{references}
\clearpage

\widetext

\begin{center}
\textbf{\large Supporting Information}
\end{center}

\setcounter{page}{1}
\setcounter{equation}{0}
\setcounter{figure}{0}
\setcounter{table}{0}
\setcounter{section}{0}
\renewcommand{\thepage}{s\arabic{page}}
\renewcommand{\theequation}{S\arabic{equation}}
\renewcommand{\thetable}{S\arabic{table}}
\renewcommand{\thefigure}{S\arabic{figure}}

\section{Comparison of the dielectric functions of gold reported in literature}\label{sec:S1}
\begin{figure}[htb!]
\centering\includegraphics{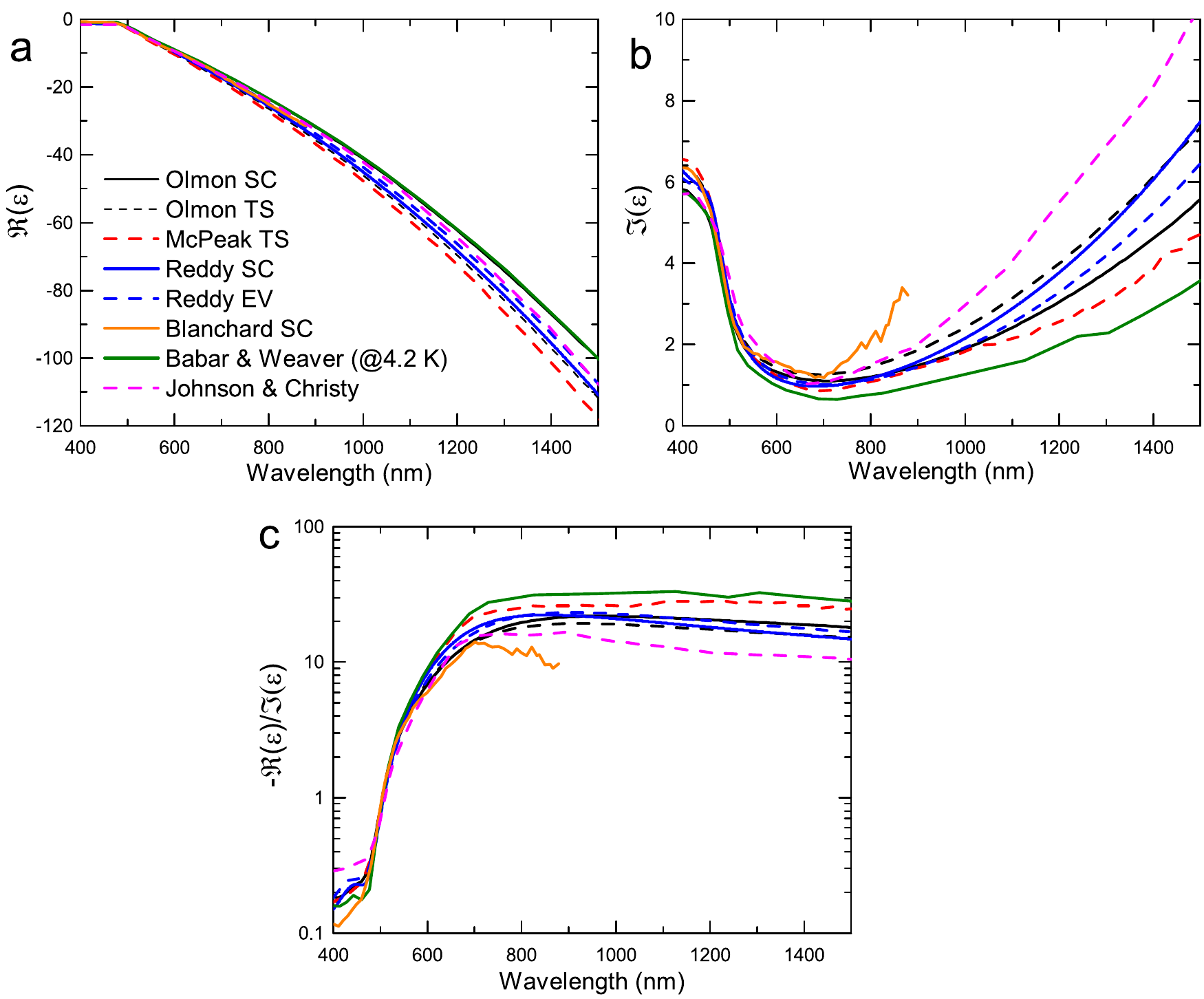}
  \caption{(a,b) Permittivity of gold (real $\Re(\varepsilon)$ and imaginary $\Im(\varepsilon)$ parts) for evaporated polycrystalline (EV), template-stripped (TS), and single-crystalline (SC) gold samples. The data is from Olmon et al. \cite{Olmon}, McPeak et al. \cite{Norris}, Reddy et al. \cite{Shalaev} (first cycle at \SI{23}{\celsius}), Blanchard et al. \cite{Blanchard}, Babar \& Weaver \cite{Babar} (measured at cryogenic temperature of \SI{4.2}{\kelvin}), and Johnson \& Christy \cite{JC}. (c) Figure-of-merit calculated as the ratio of the real to the imaginary part of the relative permittivity.}
	\label{fig:S1}
\end{figure}

\newpage
\section{Fabrication of gold flakes}\label{sec:S2}
The gold flakes were directly grown on a BK-7 glass substrate, following a modified Brust-Schiffrin method for the colloidal particle synthesis \cite{Brust-Schiffrin, Radha}. In this procedure, a solution of chloroauric acid (\ch{HAuCl4}) and water is mixed with a solution of toluene and tetrabutylammonium bromide (ToABr). ToABr acts as a phase transfer agent in this process and promotes the creation of a two-phase liquid-liquid solution. The (\ch{AuCl4})-ions contained in the organic phase are essentially the precursor for the gold crystal growth. The organic phase is then drop-casted on the substrate to undergo thermolysis at \SI{\sim 130}{\celsius} for \SI{\sim 24}{h}. From the resulting flakes, two large flakes without defects were chosen for the SPP measurements (one of them is shown in \figref[]{1}, also see \secref{S3}). A high-contrast SEM image of one flake revealed tiny wrinkles on its surface, confirmed by AFM measurements (see \secref{S4}), which, we believe, is due to the high pliability of gold (this flake was \SI{\sim 170}{\nano\meter} in thickness) and the not perfectly smooth substrate surface.

We used reagents purchased from Sigma Aldrich: chloroauric acid (\ch{HAuCl4}$\cdot$3\ch{H2O}), toluene and tetrabutylammonium bromide (ToABr). The exact recipe is as follows:
\begin{enumerate}
    \item Add \SI{1.2}{\gram} of ToABr to \SI{2.5}{\milli\liter} of toluene and stir until complete dissolution (\SI{\sim 20}{\minute}).
    \item Add \SI{1}{\milli\liter} of aqueous solution of \ch{HAuCl4} (\SI{0.5}{\molar} concentration) to the mixture.
    \item Stir at \SI{5000}{\rpm} for \SI{\sim 10}{\minute} and leave to rest for another \SI{10}{\minute}, allowing the aqueous and organic phases to separate.
    \item Clean the substrate in an ultrasonic bath, with acetone, IPA and water and blow-dry with nitrogen.
    \item Prebake the substrate at \SI{200}{\celsius} for \SI{\sim 5}{\minute} for dehydration.
    \item Drop cast \SI{\sim 20}{\micro\liter} of organic phase onto the substrate.
    \item Place the substrate on a hot plate at \SI{\sim 130}{\celsius} for \SI{\sim 24}{\hour}.
    \item Clean in toluene, acetone and IPA at \SI{75}{\celsius}. Blow-dry with nitrogen.
\end{enumerate}
\newpage

\section{Description of flakes}\label{sec:S3}
Two flakes were chosen for the SPP measurements, flake 1 and flake 2. The lateral dimensions of flake 1 are \SI{\sim 180}{} by \SI{\sim 150}{\square\micro\meter}, the thickness is \SI{\sim 170}{\nano\meter}. Flake 2 has lateral dimensions of \SI{\sim 100}{} by \SI{\sim 100}{\square\micro\meter} and a thickness of \SI{\sim 1.4}{\micro\meter}. 
\begin{figure}[htb!]
\centering\includegraphics{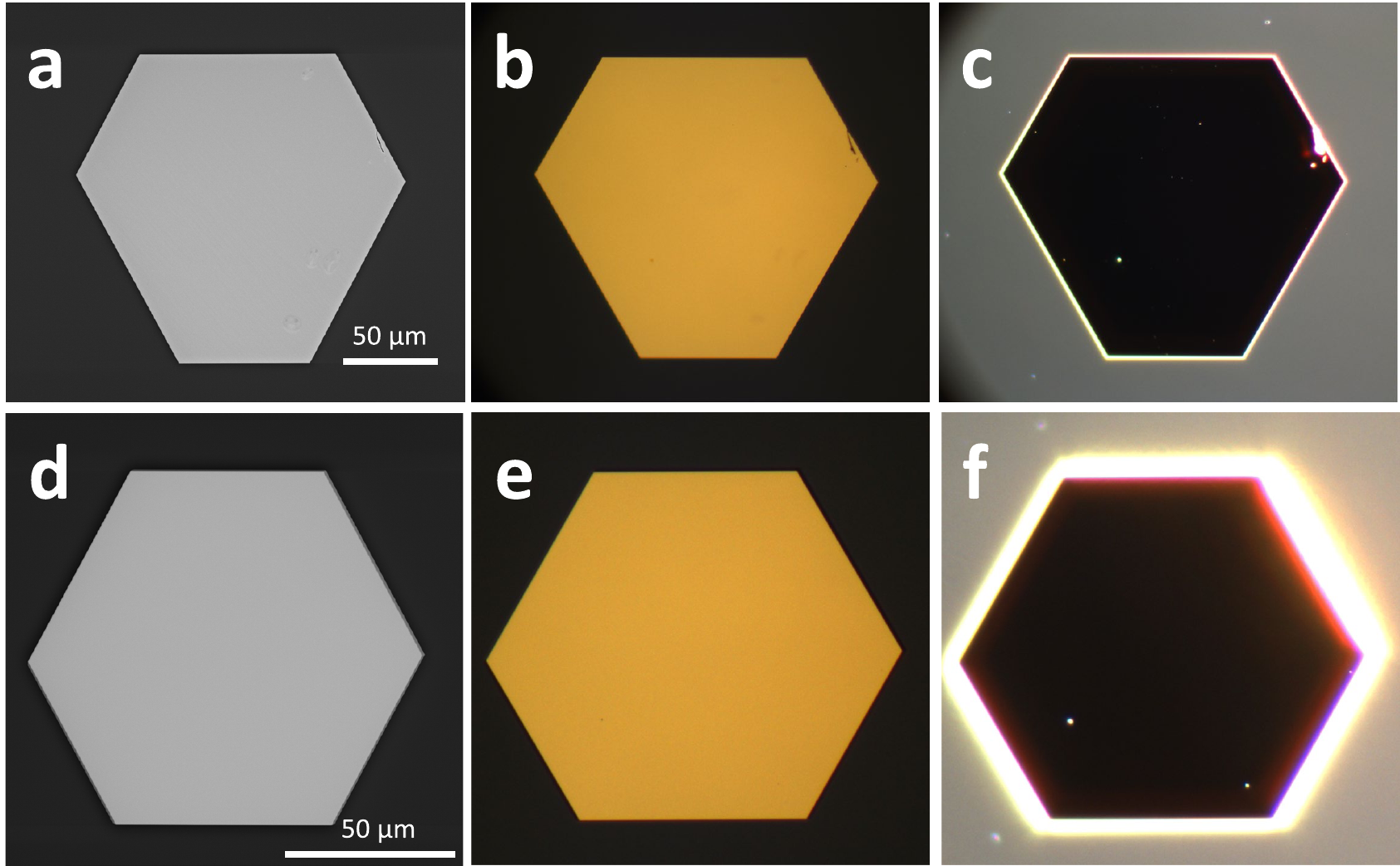}
  \caption{(a) SEM image, (b) bright-field and (c) dark-field optical microscope images of flake 1 (thickness \SI{\sim 170}{\nano\meter}), recorded with 50$\times$ objective. The bright background in the dark-field image is due to the aluminum sample holder beneath the sample. (d-f)~Same as (a-c), but of the second flake (thickness \SI{\sim 1.4}{\micro\meter}). Note the different scalebar. In (c) and (f), the edges of the flakes are overexposed to obtain better contrast in the dark regions of the flake surfaces.}
  \label{fig:S2}
\end{figure}
\newpage

\section{Flake wrinkles}\label{sec:S4}
\begin{figure}[h!]
\centering\includegraphics{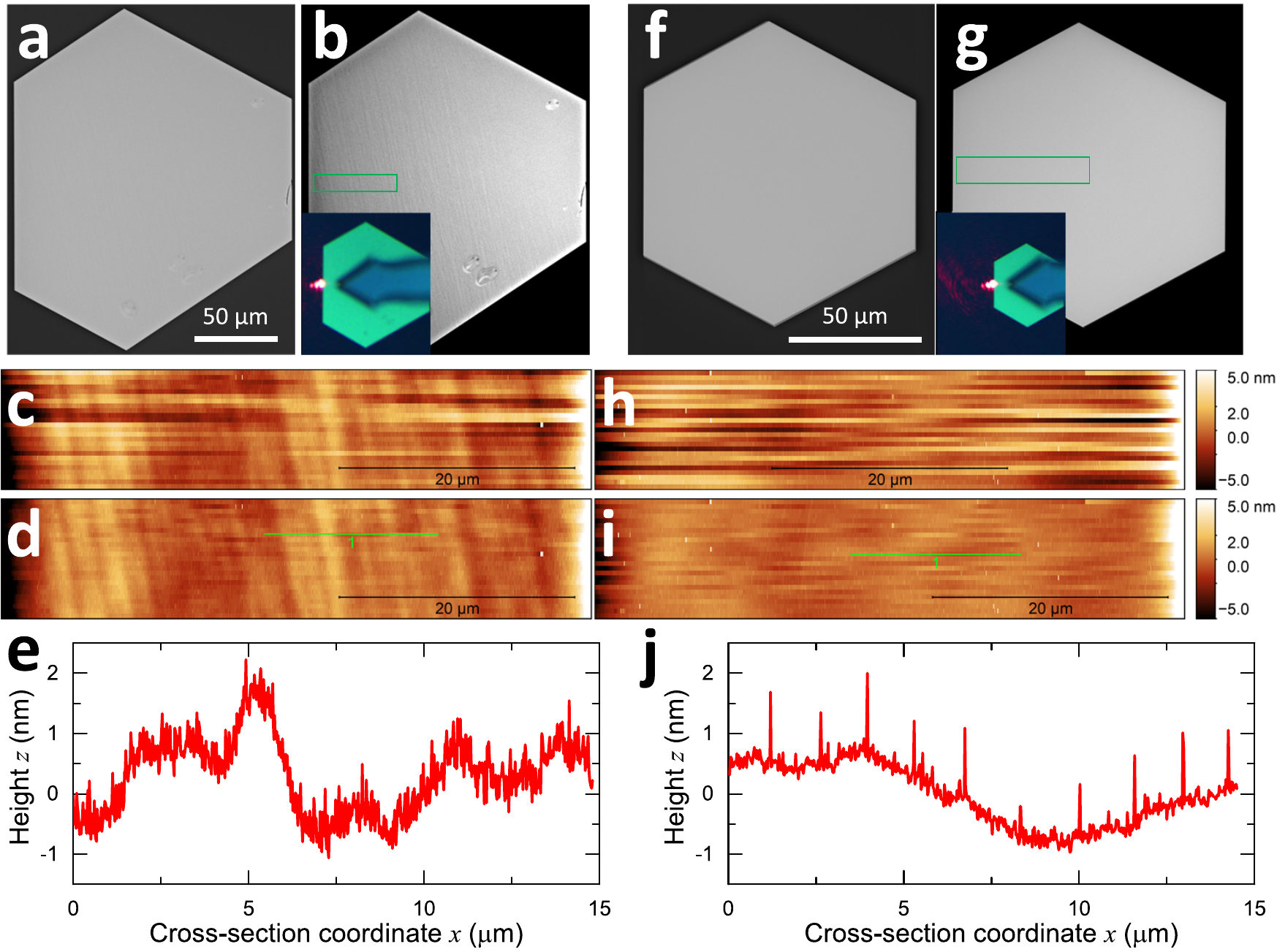}
  \caption{(a,b)~SEM images of flake 1 (thickness \SI{\sim 170}{\nano\meter}), where brightness and contrast were automatically adjusted to the dynamic range of (a)~both flake and substrate and (b)~flake only, revealing wrinkles and defects. The inset in (b)~is the optical microscope view during near-field measurements (at \SI{632.8}{\nano\meter}), with a green rectangle indicating the scan area of \SI{50}{} by \SI{10}{\square\micro\meter} (\SI{2500}{} by \SI{25}{\square\pixel}) on the SEM image. (c-d)~AFM topography, recorded during near-field mapping, after (c)~polynomial background and line-shift corrections $z(x,y)\rightarrow a(y)+b_1x+b_2 x^2+b_3 x^3+b_4 x^4$ and (d)~polynomial line-shift corrections $z(x,y)\rightarrow a(y)+b_1(y)x+b_2(y)x^2+b_3(y)x^3+b_4(y)x^4$. (e)~Line profile, indicated with a green line in (d), shows valleys of \SI{\sim 1}{\nano\meter} depth and \SI{\sim 1}{\micro\meter} width, which should have a negligible influence on SPP propagation, but might be important for applications with heterostructures, where the gold flake is expected to provide an atomically flat surface \cite{Sergio_GSP, Zayats}. (f-j)~Same as (a-e), but for the second flake (thickness \SI{\sim 1.4}{\micro\meter}), where no wrinkles were observed. The quasi-periodic jumps in (j) are due to the Stick-Slip drive technology of the SmarAct stage that moves the bottom parabolic mirror (which are less pronounced in (e) due to the larger noise). Further SEM investigations of other flakes (not shown here) supported a positive correlation between the flake thickness (estimated by the brightness in SEM image) and the presence of wrinkles.}
	\label{fig:S3}
\end{figure}

\newpage
\section{Mismatch between focused laser spot and sample displacement during the scan}\label{sec:S5}
To evaluate the lateral mismatch between the illumination laser spot position and the sample, we recorded optical microscope images from the SNOM camera of a sample with a well-defined feature (a corner of a gold square) together with a red laser focal spot. In a synchronized movement with the bottom parabolic mirror\textemdash as applied in our near-field scans\textemdash the sample was positioned at different points within the range of the piezo stage (\SI{100}{\micro\meter} in $x$- and $y$-direction). At each point, the position of the laser spot relative to the sample was measured, and it showed variations of about \SI{1}{\micro\meter} (error \SI{\sim 0.3}{\micro\meter}) (see \figref[]{S4} and \tabref{S1}). This lateral mismatch was however not accumulating over time.
\begin{figure}[htb!]
\centering\includegraphics{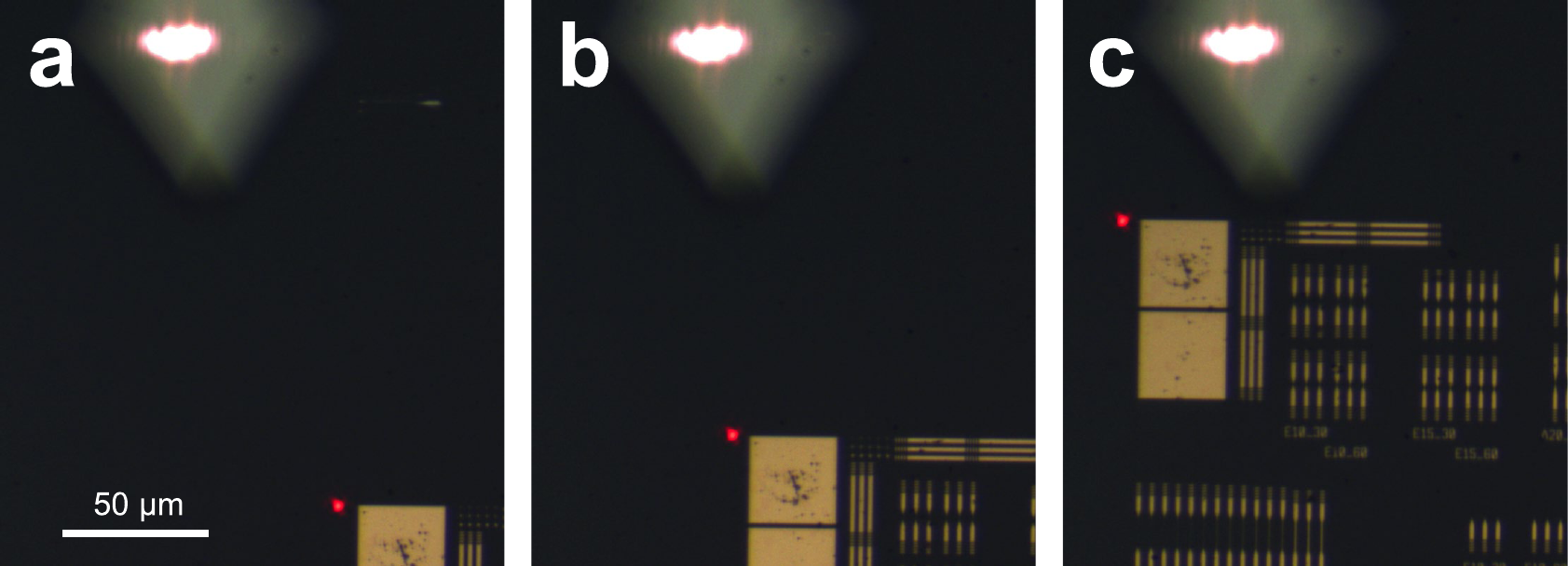}
  \caption{Optical microscope images from the SNOM camera, with both the sample and the bottom parabolic mirror positioned at [$x$, $y$] of (a)~$[1, 1]$\,µm, (b)~$[50, 25]$\,µm, and (c)~$[99, 99]$\,µm.}
	\label{fig:S4}
\end{figure}
\begin{table}[htb!]
    \centering
    \begin{tabular}{ |l|l|l|l|l|l| } 
\hline
 & $x=\SI{1}{\micro\meter}$ & $x=\SI{25}{\micro\meter}$ & $x=\SI{50}{\micro\meter}$ & $x=\SI{75}{\micro\meter}$ & $x=\SI{99}{\micro\meter}$ \\
\hline
$y=\SI{1}{\micro\meter}$ & $[-0.9, 0.1]\,\SI{}{\micro\meter}$ & $[0.2, -0.6]\,\SI{}{\micro\meter}$ & $[0.2, -0.7]\,\SI{}{\micro\meter}$ & $[0.3, -0.7]\,\SI{}{\micro\meter}$ & $[0.1, -0.9]\,\SI{}{\micro\meter}$ \\
$y=\SI{25}{\micro\meter}$ & $[-0.9, 0.4]\,\SI{}{\micro\meter}$ & $[0.3, -0.2]\,\SI{}{\micro\meter}$ & $[0.1, -0.4]\,\SI{}{\micro\meter}$ & $[0.6, -0.3]\,\SI{}{\micro\meter}$ & $[0.1, -0.5]\,\SI{}{\micro\meter}$ \\
$y=\SI{50}{\micro\meter}$ & $[-0.8, 0.6]\,\SI{}{\micro\meter}$ & $[0.5, -0.1]\,\SI{}{\micro\meter}$ & $[-0.5, 0.1]\,\SI{}{\micro\meter}$ & $[0.8, -0.4]\,\SI{}{\micro\meter}$ & $[0.1, -0.5]\,\SI{}{\micro\meter}$ \\
$y=\SI{75}{\micro\meter}$ & $[-0.8, 1]\,\SI{}{\micro\meter}$ & $[0.5, 0.4]\,\SI{}{\micro\meter}$ & $[0.1, 0]\,\SI{}{\micro\meter}$ & $[0.8, 0.2]\,\SI{}{\micro\meter}$ & $[0.1, 0]\,\SI{}{\micro\meter}$\\
$y=\SI{99}{\micro\meter}$ & $[-1.1, 1.2]\,\SI{}{\micro\meter}$ & $[0, 0.5]\,\SI{}{\micro\meter}$ & $[-0.2, 0.4]\,\SI{}{\micro\meter}$ & $[0.3, 0.3]\,\SI{}{\micro\meter}$ & $[-0.2, 0.3]\,\SI{}{\micro\meter}$ \\
\hline
\end{tabular}
    \caption{Mismatch between focused laser spot and sample position $[\Delta x, \Delta y]$, measured for different sample coordinates (see row and column headers).}
    \label{tab:S1}
\end{table}

\newpage
\section{SPP Gaussian beam and laser alignment}\label{sec:S6}
When a well aligned free-space 3D Gaussian beam is incident on the flake edge, it will excite SPPs with near-perfect 2D Gaussian beam shape, propagating along the gold surface and perpendicularly to the flake edge. The field of a two-dimensional SPP Gaussian beam propagating along the $x$-axis, with its focal point at the origin and beam waist of $w_0$, can be written as:
\begin{equation}\label{eq:S1}
    E(x,y) =    E_0 \sqrt{\frac{w_0}{w(x)}} \exp{\left[\frac{-y^2}{w^2(x)}\right]} \exp{\left[ik_0 N_{\mathrm{SPP}}\frac{y^2}{2R(x)}\right]}\exp{\left[i\Phi(x)\right]}\exp{\left[ik_0N_{\mathrm{SPP}}x\right]}\,,
\end{equation}
where $w(x)=w_0\sqrt{(1+(x/x_{\mathrm{R}})^2)}$ is the increasing beam width, $R(x)=x[1+(x/x_{\mathrm{R}})^2]$ is the radius of curvature of the circular beam wavefront, $\Phi(x)=\frac{1}{2}\arctan{(x/x_{\mathrm{R}})}$ is the Guoy phase shift (longitudinal phase shift upon passing through the focal point), and $x_{\mathrm{R}}=(\pi w_0^2)/\lambda_{\mathrm{SPP}}$ is the Rayleigh range. After performing Fourier filtering of the recorded near-field maps, we fit each cross-section $E_{\mathrm{f}}(y)$ along the propagation direction with a Gaussian to find the SPP beam width as a function of the propagation coordinate and fit it with $w(x)=w_0\sqrt{1+\frac{(x-x_0)^2}{x_{\mathrm{R}}^2}}$ to determine the SPP beam waist $w_0$, its position $x_0$ and the Rayleigh range $x_{\mathrm{R}}$. The obtained values are listed in \tabref{S2}. Due to the relatively large Rayleigh ranges, the influence of the SPP beam divergence, i.e., the third term in \equref{S1} ($\exp{\left[ik_0 N_{\mathrm{SPP}}\frac{y^2}{2R(x)}\right]}$), becomes negligible. Then, only the first two terms describe the amplitude variation due to the beam divergence. Thus, integrating of $|E_{\mathrm{f}}|^2$ for each cross-section compensates for the beam divergence and provides a decay that is solely SPP absorption related: $\int E_{\mathrm{f}}^2 dy\propto|E_0|^2\exp{\left[-2k_0\Im(N_{\mathrm{SPP}})x\right]}$ (note that $L_{\mathrm{p}}=\frac{\lambda_0}{4\pi\Im(N_{\mathrm{SPP}})}$ and $\exp{\left[-2k_0\Im(N_{\mathrm{SPP}})x\right]}=\exp{\left[-\frac{x}{L_{\mathrm{p}}}\right]}$). The gradient of the Gaussian beam phase along the $x$-axis is $\frac{1}{2x_{\mathrm{R}}}[1+(\frac{x}{x_{\mathrm{R}}})^2]^{-1}+k_0\Re(N_{\mathrm{SPP}})$, where the first term originates from the Gouy phase shift. However, it is negligibly small for our estimated Rayleigh ranges, therefore the phase gradient can be used to directly find the real part of the effective mode index.

From the above description, it follows that the beam waist $w_0$ should be large, so that the Rayleigh range $x_{\mathrm{R}}$ is of the order of the scan length to avoid large correction factors due to the beam divergence. To accurately determine the SPP propagation length $L_{\mathrm{p}}$, the scan length $L_{\mathrm{s}}$ should further be sufficiently large ($L_{\mathrm{s}}>L_{\mathrm{p}}$). At the same time, the correction of the beam divergence by integrating $\int |E_{\mathrm{f}}|^2 dy$ will only work if the scan width $w_{\mathrm{s}}$ is also large, so that the finite integral within the scan width is close to the infinite integral (the difference between finite and infinite integrals, which should be much smaller than the SPP propagation losses, is shown in \figref[]{S5}). Therefore, the following relations were used to roughly define the incident beam size and the scan area, using the expected SPP propagation length:
\begin{equation*}
\begin{split}
    &L_{\mathrm{p}}\leq L_{\mathrm{s}} \leq x_{\mathrm{R}}, \text{with } x_{\mathrm{R}}=\frac{\pi w_0^2}{\lambda_{\mathrm{SPP}}}\,,\\
    &w_0,w(x=L_{\mathrm{s}})\ll\frac{1}{2}w_{\mathrm{s}}\,.
    \end{split}
\end{equation*}
Since the focus of the incident laser beam was aligned at the flake edge, the smallest width of the SPP Gaussian beam is expected at the beginning side of the scan.
\begin{figure}[htb!]
\centering\includegraphics{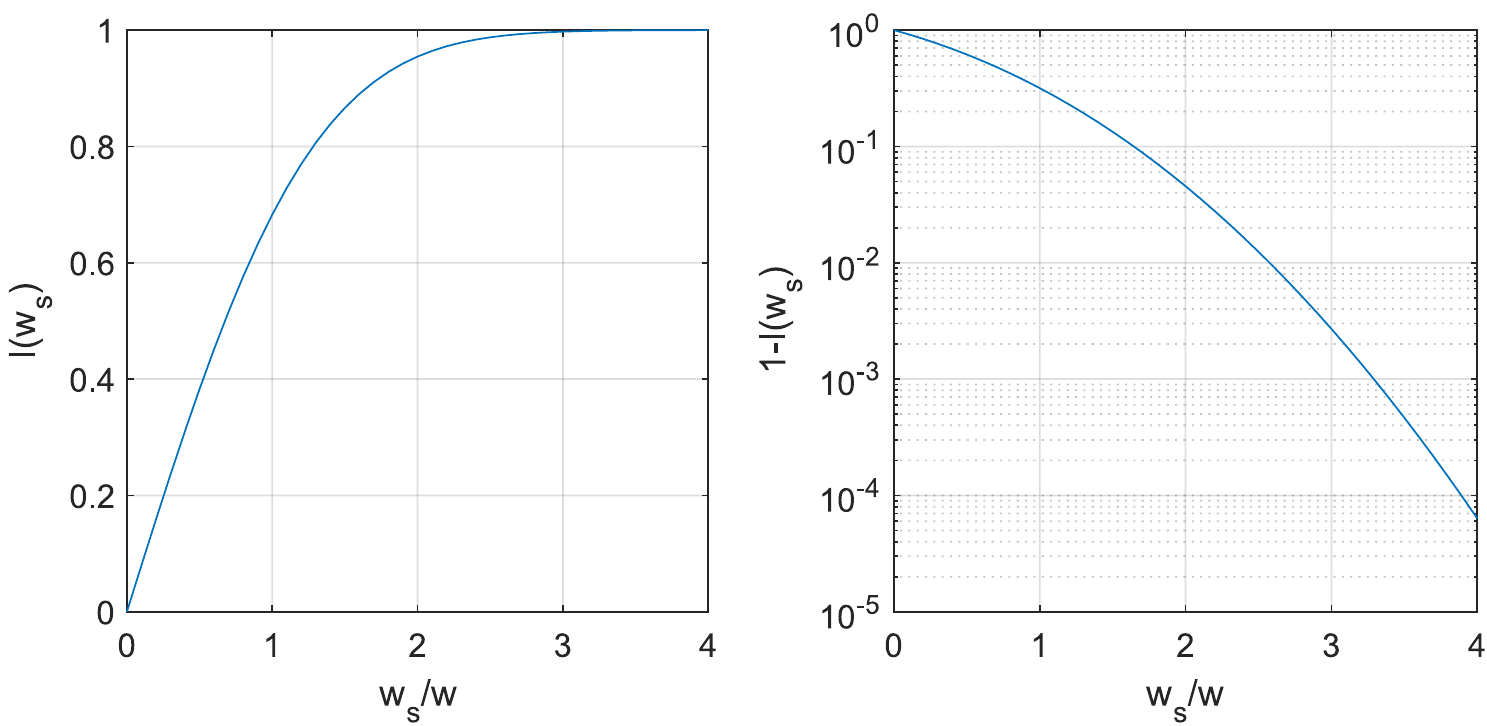}
  \caption{Difference between finite and infinite integrals of $\int |E|^2 dy$ as the function of limits, where $I(w_{\mathrm{s}})=\int_{-0.5w_{\mathrm{s}}}^{0.5w_{\mathrm{s}}} \exp{(-2\frac{y^2}{w^2})} dy / \int_{-\infty}^{\infty}\exp{(-2\frac{y^2}{w^2})} dy$.}
	\label{fig:S5}
\end{figure}
\begin{table}[htb!]
    \centering
    \begin{tabular}{ |l|l|l|l|l|l| } 
\hline
Wavelength $\lambda_0$ / \SI{}{\nano\meter} & 532 & 594 & 632.8 & 729 & 800 \\
\hline
$\lambda_{\mathrm{SPP}}$ / \SI{}{\nano\meter} & 493 & 566 & 627 & 715 & 784 \\ 
$L_{\mathrm{p}}$ / \SI{}{\micro\meter} & 0.99 & 6.2 & 13 & 43 & 80 \\ 
SPP beam waist $w_0$ / \SI{}{\micro\meter} & 1 & 3.4 & 3 & 5.5& 5.7 \\ 
SPP beam waist at end of scan $w(x=L_{\mathrm{s}})$ / \SI{}{\micro\meter} & 1.6 and 1.8 & 3.6 & 4.5 & 5.9 & 6.1 \\
Rayleigh range $x_{\mathrm{R}}$ / \SI{}{\micro\meter} & 6 & 63& 46& 131& 127 \\
Scan size $L_{\mathrm{s}} \times w_{\mathrm{s}}$ / \SI{}{\square\micro\meter} & $10\times 10$ and $8\times 5$ & $20\times 10$ & $50\times 10$ & $50\times 20$ & $50\times 20$ \\
\hline
\end{tabular}
    \caption{Estimated SPP Gaussian beam parameters.}
    \label{tab:S2}
\end{table}

\clearpage
\section{Influence of Fourier filtering and the choice between 3rd and 4th harmonic near-field map}\label{sec:S7}
The Fourier filtering procedure we applied during analysis is demonstrated in \figref[]{S6}, using data for \SI{594}{\nano\meter} illumination wavelength as an example, recorded at the 3rd (\figref[a]{S6}) and the 4th (\figref[b]{S6}) harmonic of the tip-oscillation frequency. Left column shows the amplitude of raw near-field maps, which were Fourier-transformed along the propagation direction $x$, using the extended discrete Fourier transform (EDFT, \cite{EDFT}) of 5 times longer length than in the original data (to increase the resolution). The average spectra of these Fourier maps (averaged along $y$-axis) are shown in \figref[d]{S6}. Next to the main SPP-related peak at $k_x/k_0\approx 1.05$, to which the spectra are normalized, the Fourier spectra show a higher frequency contribution around $k_x/k_0\approx 1.3$ (see discussion in the next section).
\begin{figure}[htb!]
\centering\includegraphics{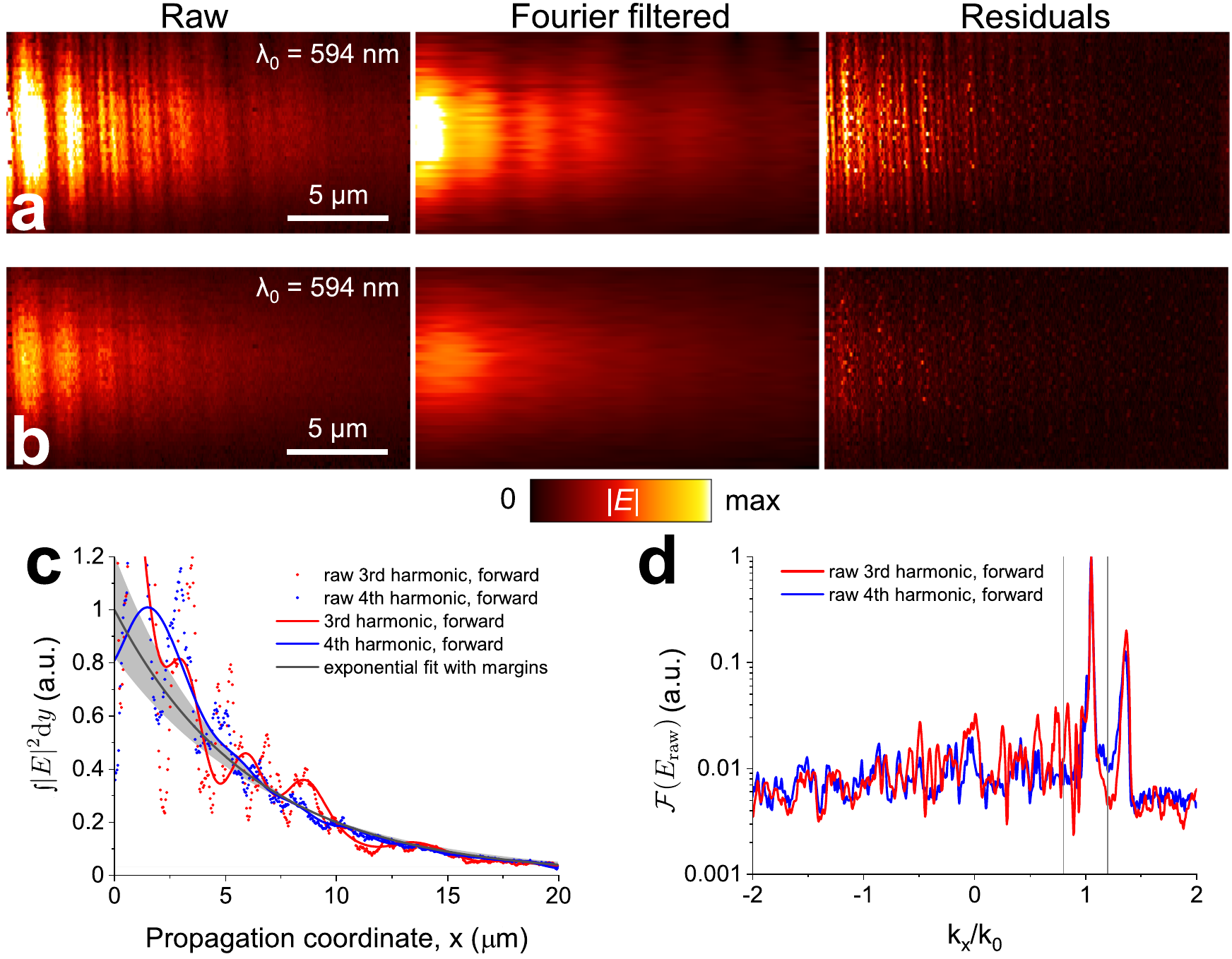}
  \caption{Near-field amplitude maps of the SPP propagation (excited at $\lambda_0=\SI{594}{\nano\meter}$), recorded at the (a)~3rd and (b)~4th harmonic of the tapping frequency. Shown are the raw and Fourier-filtered maps, as well as the map of the removed background. (c)~Decay traces corresponding to the raw and Fourier-filtered maps in (a,b). The black line with grey area corresponds to the exponential decay with the propagation length determined for the respective wavelength (Table 1 in the main text). (d)~Average Fourier spectra for (a,b), calculated in SPP propagation direction. Grey vertical lines indicate limits of the apodization function, applied to select only the contributions $0.8<k_x/k_0 <1.2$ to be used for further analysis ($0.8<k_x/k_0<1.3$ for $\lambda_0=\SI{532}{\nano\meter}$).}
	\label{fig:S6}
\end{figure}
Further, some lower frequency contributions are visible in the range $|k_x/k_0|<1$, which we attribute to the bulk scattering by the cantilever (so-called background), because they are suppressed by increasing the harmonic number (which means a weak dependance on a tip-sample distance). Fourier filtering (with a simple rectangular apodization function, $0.8<k_x/k_0<1.2$ here) largely removes them from the 4th harmonic near-field map, leaving an almost pure SPP peak (see filtered near-field map and residuals in the middle and the right column of \figref[b]{S6}, correspondingly). For the 3rd harmonic, however, there is a background peak at $k_x/k_0\approx 0.9$, which survives the filtering, resulting in the beating pattern in the filtered near-field map (the middle column of \figref[a]{S6}). Narrowing down the filtering window will help removing this contribution, but care should be taken to avoid filtering artifacts (when filtering limits are close to the SPP mode). Because of the overall lower background and good signal-to-noise ratio, we only use the 4th harmonic near-field measurements for analysis in this publication (though the values from the 3rd harmonic near-field measurements are in good agreement).

\clearpage
\section{Analysis of interference fringes in the near-field maps near the excitation edge}\label{sec:S8}
\begin{figure}[htb!]
\centering\includegraphics{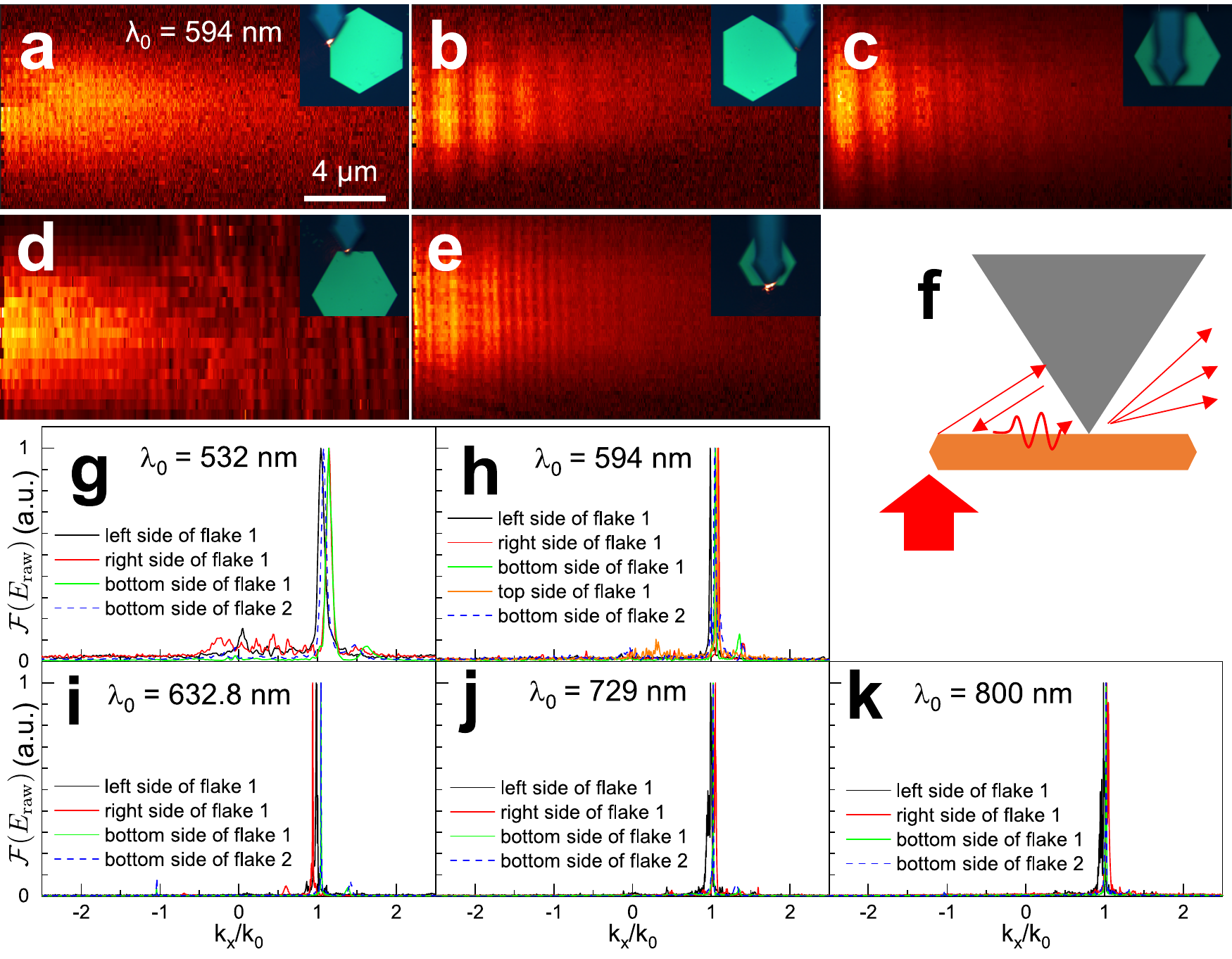}
  \caption{Near-field maps of SPPs, excited from (a-d)~different edges of flake 1 and (e)~bottom side of flake 2. Insets show screenshots from s-SNOM top-view camera. (f)~Possible explanation for the artificial mode: the diffraction at the flake edge is back-reflected by the tip’s pyramid, exciting SPPs, which propagate and are scattered by the tip. Due to the divergence of the diffracted light, this artificial mode is only pronounced near the edge. In the case of the excitation from the top edge (d), there is a strong back-reflection by the cantilever (when the tip is above the flake and far from the edge), which results in strong background in the near-field map. Therefore, the measurements in this excitation configuration were not used in the final analysis. (g-k)~Average Fourier spectra in the propagation direction for indicated wavelengths. The position of the SPP peak is not the same for different excitation configurations because the incident beam was not perfectly normal to the sample (the shown mismatch in $k_x/k_0=0.05$ corresponds to the angle of \SI{\sim 3}{\degree}). Note the position of the artificial mode varies with the wavelength and the excitation configuration, making the explanation in (f) just an assumption.}
	\label{fig:S7}
\end{figure}

\clearpage
\section{Cantilever-flake configurations and fitting intervals}\label{sec:S9}
\begin{figure}[htb!]
\centering\includegraphics{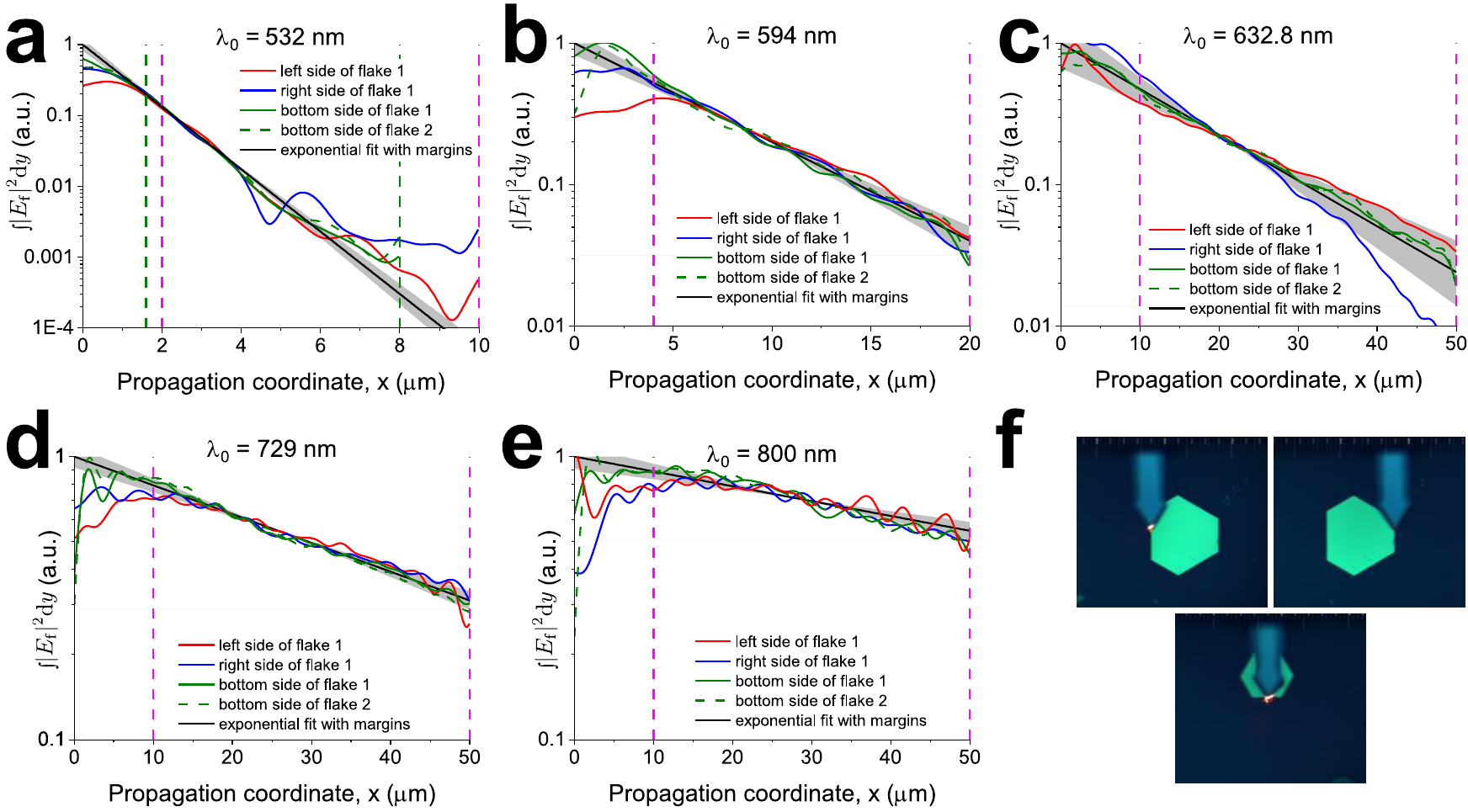}
  \caption{To suppress the influence of the sample and the setup (asymmetry of the tip, orientation of the top parabolic mirror collecting the tip’s scattering), and for error estimation, we scanned with different orientations of the cantilever to the flake edge. (a-e)~Decay traces obtained from Fourier-filtered near-field maps for the different cantilever-flake edge orientations, plotted for each wavelength. Fitting was performed within the magenta markers (neglecting the first 20\% of propagated distance; the fitting interval of the shorter bottom-side scans in (a) is marked in green). The black line with grey area corresponds to the exponential decay with the propagation length determined for the respective wavelength (Table 1 in the main text). (f)~The cantilever-flake edge orientations, viewed through the optical microscope. From left to right: left side of flake 1, right side of flake 1, bottom side of flake 2.}
	\label{fig:S8}
\end{figure}

\end{document}